\documentclass[pra,twocolumn]{revtex4}

\usepackage[utf8]{inputenc}

\usepackage{array}
\newcolumntype{P}[1]{>{\centering\arraybackslash}p{#1}}

\usepackage{bm}
\usepackage{amsmath}

\usepackage{amssymb}
\usepackage{graphicx}
\usepackage{multirow}

\begin{document}

\newtheorem{theorem}{Theorem}
\newtheorem{corrolary}{Corollary}

\def\pr{\prime}
\def\be{\begin{equation}}
\def\en#1{\label{#1}\end{equation}}
\def\d{\dagger}
\def\bar#1{\overline #1}
\def\U{\mathcal{U}}
\newcommand{\per}{\mathrm{per}}
\newcommand{\rd}{\mathrm{d}}
\newcommand{\vare}{\varepsilon }

\newcommand{\br}{\mathbf{r}}
\newcommand{\m}{\mathbf{m}}
\newcommand{\s}{\mathbf{s}}
\newcommand{\bk}{\mathbf{k}}
\newcommand{\bl}{\mathbf{l}}
\newcommand{\cL}{\mathcal{L}}

\title{Classical sampling from  noisy Boson Sampling  and the   negative probabilities }

\author{V. S. Shchesnovich }

\affiliation{Centro de Ci\^encias Naturais e Humanas, Universidade Federal do
ABC, Santo Andr\'e,  SP, 09210-170 Brazil }

\begin{abstract}
It is known that,   by accounting for  the   multiboson interferences up to a finite  order, the output distribution of  noisy Boson Sampling, with   distinguishability of bosons serving  as  noise,  can be approximately  sampled from  in  a  time polynomial in the total number of bosons.  The drawback of this approach is that the  joint  probabilities of completely distinguishable bosons,  i.e., those that  do not interfere at all,   have to be  computed also.  In trying to   restore the ability to sample from the  distinguishable bosons  with computation of  only the single-boson  probabilities,      one faces   the following issue: the quantum  probability factors  in a convex-sum expression,  if truncated to a finite order  of multiboson    interference,      have, on average,  a finite amount of negativity in a random interferometer.   The truncated  distribution does become  a proper one, while   allowing   for sampling   from it  in  a polynomial time, only in a  vanishing domain close to the completely distinguishable bosons.   Nevertheless, the conclusion  that   the negativity issue is  inherent  to all  efficient classical  approximations to noisy Boson Sampling may be  premature. I outline the direction for a whole new program, which seem to  point to a   solution.   However its success depends on   the asymptotic behavior    of  the  symmetric  group characters,  which is   not known.       \end{abstract}
\maketitle

\section{Introduction}

Boson Sampling model  \cite{AA}  is one of the proposals for near term quantum advantage with intermediate size quantum systems \cite{QSup}, with the advantage that it does not  involve interactions between  quantum subsystems   (individual bosons) with the  promised quantum advantage over classical simulations coming solely from the  Bose-Einstein  statistics.  On the experimental side, the photons are quite suitable source of  non-interacting bosons and     Boson Sampling with $N=20$ photons was recently demonstrated experimentally \cite{20Ph60M}, still short of the believed threshold  $N\approx  50$ bosons \cite{QSBS,Cliffords}  for the advantage over  digital computers.  Instead the focus shifted to the so-called Gaussian Boson Sampling  \cite{GBS1,GBS2} with the squeezed states of light at input,  which admits much  better scalability  in experiments \cite{ExpGBS2}.    On the other hand, one has also to keep in mind the fact   that    realistic sources  and other setup parts feature  some amount of  noise.     Such are realistic photon sources, producing only     partially indistinguishable photons,  due to imperfect internal state matching or the optical path mismatch  in propagation in a realistic interferometer.  This and other sources of noise    may  severely affect the  possible quantum advantage  by allowing an approximate efficient classical sampling.    In this respect the   single-photon Boson Sampling  model   allows for an  analytical analysis  of how  the quantum advantage  is affected by the inevitable experimental noise due  to uncontrolled  partial  distinguishability of photons.  There is the classical limit, the    completely distinguishable photons.  In Ref.  \cite{R1} it was shown that by  employing a cut-off on the higher-orders of multi-photon interferences  and simulating classically the resulting approximate  model    one can efficiently  sample classically, to a small $N$-independent error, from such a noisy Boson Sampling.  Similar approach can be employed for other noise sources,  such as noise in interferometer  \cite{LP,KK,Arkh} due to the    equivalence links  between different noise models \cite{VS2019}.  The approach of Ref. \cite{R1}    requires  efficient computation of the joint transition  probabilities of completely distinguishable bosons    by employing the JSV algorithm \cite{JSV}.    However,  this is a  strange feature of  a \textit{sampling} algorithm, since completely  distinguishable bosons behave as classical particles: they can  be sampled from in linear time, e.g., by sending particles one by one through the interferometer. 

To investigate whether one can do better that the  algorithm of Ref. \cite{R1}, especially  in dealing with the classical particles,    is the main objective  of the present work. One would expect, as a better algorithm,  an algorithm  which is polynomial in the total number of bosons  for any finite value of the distinguishability parameter and which  does not rely on \textit{computing} the probabilities of  joint  transitions  of the completely distinguishable bosons (classical particles).  An algorithm which satisfies the second condition was  presented  in Ref. \cite{Moy}, however,  it cannot be a  polynomial algorithm  in the total number of bosons  for a   finite distinguishability parameter. 

The text is organized as follows. In the next section, section \ref{sec2}, I  summarize  the appropriate     description of  partially distinguishable bosons    \cite{VS2014,PartDist}, and formulate the condition for  the  partial  distinguishability function  for the proper   probability distribution. In section \ref{sec3} I recall the approach    of   Ref. \cite{R1}  of simulating Boson Sampling with multi-boson interferences up to a fixed order $R$ and the reasons  why it  works.  In section   \ref{sec5},   the    quantum probability factor truncated to lower-order interference terms is shown to have  some    finite negativity, thus  preventing direct sampling from  the  convex-sum expansion for probability.   In section \ref{sec4},    it is shown that   the  approach    of Ref. \cite{R1}  produces  a proper probability distribution if  the distinguishability parameter  satisfies $x\le \frac{1}{N-K}$ for  a free parameter $K$, at a heavy  price of the sampling complexity  scaling as  $O(K^2 4^K)$. Finally,  in section \ref{sec6} I outline the direction for a new   program, which might permit  one to  find  proper       approximating  distributions for the  noisy Boson Sampling distribution. Section \ref{sec6} contains a  short summary of the results. 

\section{Models of distinguishability  with   proper probability distributions   }
\label{sec2}
 
 Boson Sampling model performs unitary transformation  $U$ of   size $M\times M$   on  the Fock state of   $N$ indistinguishable bosons in different input ports, say $k=1,\ldots, N$ to an output Fock state  in the so-called no-collision regime, when  all bosons end up in different output ports $l_1,\ldots, l_N$ (the odds to have this in a random multiport is estimated to be  $1-O(N^2/M)$ \cite{AA}).   The output probability is  given by  the product of the quantum amplitude of the transition and its complex conjugate,
\begin{eqnarray}
\label{Eq1}
&& p(l_1,\ldots,l_N)  = \left|\sum_{\sigma} \prod_{k=1}^N U_{\sigma(k),l_k}\right|^2\nonumber\\
&&  =  |\per U[1,\ldots,N|l_1,\ldots,l_N]|^2,
\end{eqnarray}
where  the amplitude  is given by the  matrix permanent \cite{Minc}, i.e.,   the summation is over all permutations $\sigma$ of $N$ objects  (a.k.a. the symmetric group $S_N$).

Realistic  description of a physical setup (e.g., with photons) must  account for distinguishability of  bosons. This  can be  done by introducing a  single  function $J(\sigma)$ on the symmetric group  $S_N$, called the distinguishability function, which weights the  path-dependent  interferences  of $N$ bosons   in a quantum amplitude and its conjugate  for different  permutations $\sigma$, i.e., the expression in Eq. (\ref{Eq1}) generalizes  \cite{VS2014,PartDist} as follows
\be
p^{(J)}(l_1,\ldots,l_N) =  \sum_{\sigma_1,\sigma_2} J(\sigma_1\sigma_2^{-1})\prod_{k=1}^N U^*_{\sigma_1(k),l_k}U_{\sigma_2(k),l_k}.
\en{Eq2}  
The distinguishability function $J(\sigma)$  reflects the internal states of bosons, described by  the    density matrix $\hat{\rho}^{(N)}\in  H^{\otimes N}$,  in the tensor product of   Hilbert spaces $H$ of each boson,   and is given by the trace-product with  the unitary  representation  $\hat{P}_\sigma$  of   $\sigma$  in  $H^{\otimes N}$:
  \be
  J(\sigma) = \mathrm{tr}\left( \hat{P}_\sigma \hat{\rho}^{(N)} \right).
  \en{Eq3}
  When the internal state is completely symmetric, i.e. $\hat{P}_\sigma \hat{\rho}^{(N)} =   \hat{\rho}^{(N)} $, the bosons in the  $\hat{\rho}$ are completely indistinguishable \cite{VS2014,PartDist},  and  we get  the probability as in Eq. (\ref{Eq1}); the completely distinguishable bosons correspond to $J(\sigma) = \delta_{\sigma,I}$, where $I$ is the trivial permutation,  with the probability in Eq. (\ref{Eq2})  being  equal to the matrix permanent of a  positive matrix  with elements $|U_{kl}|^2$. 

The $J$-function of Eq. (\ref{Eq3})  is positive definite, i.e., for  an arbitrary function $Z_{\sigma}$ on $S_N$  we have 
\be
\sum_{\sigma_1,\sigma_2} J(\sigma_1\sigma_2^{-1})Z_{\sigma_1}Z^*_{\sigma_2} \ge 0. 
\en{Eq4} 
Observe that Eq. (\ref{Eq4}) presupposes that  $J^*(\sigma) = J(\sigma^{-1})$ readily satisfied by  $J(\sigma)$ of Eq. (\ref{Eq3}) due to the unitarity of the symmetric group representation, $P^\dag_\sigma = P_{\sigma^{-1}}$.  The group property  $\hat{P}_{\sigma_1\sigma^{-1}_2} = \hat{P}_{\sigma^{-1}_1}\hat{P}{\sigma_2}$  guarantees   the positive definiteness of $J$ in Eq. (\ref{Eq3}):
\be
 \mathrm{tr}\left( \sum_{\sigma_1}Z_{\sigma_1}\hat{P}_{\sigma_1}  \hat{\rho}^{(N)}  \sum_{\sigma_2}Z^*_{\sigma_2} \hat{P}^\dag_{\sigma_2}\right)  \ge 0,
\en{Eq5} 
since the density matrix of the internal state  of bosons $\hat{\rho}$ is a positive definite operator in $H^{\otimes N}$.  Another important property is  the normalization   $J(I) =1$  (i.e., the sum of probabilities  must be  equal to $1$).  These two  properties guarantee that the probabilities in Eq. (\ref{Eq2}) constitute a proper distribution. It has been shown in Ref. \cite{PRL2016} that all normalized positive definite functions on the symmetric group  $S_N$ can be represented in the form of Eq. (\ref{Eq3}) with some (possibly  entangled) mixed state $\hat{\rho}\in H^{\otimes N}$ of $N$ single bosons. Thus  one can on choice   work  either with the internal state $\hat{\rho}$ or with the distinguishability function description  in dealing with partially distinguishable bosons. 

 On the other hand performing an arbitrary   \textit{approximation} in the expression for the proper probability distribution may result in a non-proper one.  
The    model of  Ref. \cite{R1} is   obtained by     considering  that  bosons are  in some pure  internal states  having a uniform cross-state overlap $x$, or, alternatively, considering  that the  boson at input port  $k$ is   in the following  mixed state 
\be
\hat{\rho}_k = x|\phi_0\rangle\langle\phi_0| +(1-x) |\phi_k\rangle\langle\phi_k|, \quad  \langle\phi_i|\phi_j\rangle = \delta_{ij},  
\en{Eq6} 
in this case $\hat{\rho}^{(N)} = \hat{\rho}_1\otimes \ldots \otimes \hat{\rho}_N$  (see also Ref. \cite{VS2019}).  A  non-proper approximation to the proper distribution is obtained  in Ref. \cite{R1}  by  imposing a cut-off  on  the  two  sums in Eq. (\ref{Eq2}),  such that the minimum  number of fixed points in the relative permutation (i.e., the number of $1$-cycles  $C_1$) is bounded from below:  $C_1(\sigma_1\sigma^{-1}_2) \ge N-R$ for some fixed ($N$-independent) $R$. The  distinguishability function  $J_N(\sigma) = x^{N-C_1(\sigma)}$ of  the model in Eq. (\ref{Eq6})  is  replaced    accordingly by the following  function 
\be
 J_R(\sigma) = \sum_{m=N-R}^N x^{N-m} \delta_{C_1(\sigma),m},
\en{Eq7}
 which does not satisfy the positivity property of Eq. (\ref{Eq4}). To see this and the underlying reason for lost positivity, let us formulate more general condition of positivity valid for  all models of   similar type, i.e., 
\be
F(\sigma) =    \sum_{m=0}^N a_m \delta_{C_1(\sigma),m},
\en{Eq8}
where $a_m$ are real parameters satisfying the normalization condition $F(I)=1$, i.e., $a_N=1$. To this goal we expand $\delta_{C_1(\sigma),m}$  over the partitions of the  set $1,\ldots, N$  into the variable set of $m$ fixed points $\bk^{(m)} = (k_1,\ldots k_m)$ and its complement (i.e., the set of derangements). Then   
\begin{eqnarray}
\label{Eq9}
&& \delta_{C_1(\sigma),m} = \sum_{\bk^{(m)}}\left( \prod_{k\in \bk^{(m)}} \delta_{\sigma(k),k}\right)\prod_{k\notin \bk^{(m)}} (1-\delta_{\sigma(k),k}) \nonumber\\
&&= \sum_{n=m}^N\binom{n}{m} (-1)^{n-m}\sum_{\bk^{(n)}} \prod_{k\in \bk^{(n)}}\delta_{\sigma(k),k}, 
\end{eqnarray} 
where we have expanded the second factor and combined the fixed points into a bigger set $\bk^{(n)}$ of size $n$.  Now we substitute the expansion of Eq. (\ref{Eq9})  into the model Eq. (\ref{Eq8})  and exchange the order of summation
\begin{eqnarray}
\label{Eq10}
 && F(\sigma) =    \sum_{m=0}^N a_m \delta_{C_1(\sigma),m} \nonumber\\
 && = \sum_{n=0}^N \sum_{m=n}^N \binom{n}{m} (-1)^{n-m}a_m \prod_{k\in \bk^{(n)}}\delta_{\sigma(k),k}.
\end{eqnarray} 
 For the function in Eq. (\ref{Eq10})  to be positive definite it is sufficient  to require   the  coefficients   at  the functions $j_{\bk^{(n)}}(\sigma) = \prod_{k\in \bk^{(n)}}\delta_{\sigma(k),k}$ to be  positive, since  $j_{\bk^{(n)}}(\sigma)$ is a positive definite function, e.g., for $\bk^{(n)} = (1,\ldots,n)\equiv [n]$ we have the following   representation 
 \be
 j_{[n]}(\sigma) = \mathrm{Tr}\left\{ \hat{P}_\sigma \prod_{k=1}^n{}^{\otimes} |\phi_k\rangle\langle\phi_k| \otimes \left(|\phi_0\rangle\langle\phi_0|\right)^{\otimes N-n}  \right\},
 \en{Eq11}
with $\langle\phi_i|\phi_j\rangle=\delta_{ij}$, which is  obviously positive definite. Setting the coefficients  at the positive definite functions $j_{\bk^{(n)}}(\sigma)$  in Eq. (\ref{Eq10}) to be
\be
 b_n \equiv  \sum_{m=0}^n \binom{n}{m}(-1)^{n-m}a_m
\en{Eq12A}
and inverting  the   summation in their definition   we get the   conditions   for the positivity of the function $F(\sigma)$  in Eq. (\ref{Eq10}):
\be
a_m = \sum_{n=0}^m \binom{m}{n}b_n,\quad b_n\ge 0. 
\en{Eq12B}
 Therefore, among  the   functions  $F(\sigma)$ in  the form given by Eq. (\ref{Eq10}) the  functions 
\be
F^{(+)}(\sigma) = \sum_{n=0}^N b_n  \sum_{m=n}^N \binom{m}{n}  \delta_{C_1(\sigma),m},\quad b_n\ge 0 
\en{Eq13} 
  are positive definite functions on $S_N$. 
 
 The  partial distinguishability  model   of  Ref. \cite{R1}, i.e., $J_N(\sigma)=x^{N-C_1(\sigma)}$, can be also cast in the form of Eq. (\ref{Eq13})  with $b_n = x^{N-n}(1-x)^n$ (observe that    $0\le x\le 1$, see Eq. (\ref{Eq6})),  which fact  can be directly verified by  exchanging the summation and using the binomial theorem. 
 
The physical interpretation of the condition in Eq. (\ref{Eq13})  follows from  application of the expansion of Eq. (\ref{Eq9}) for  each term with positive coefficient in Eq. (\ref{Eq13}). We obtain  the relation 
\be
 \sum_{m=n}^N \binom{m}{n}  \delta_{C_1(\sigma),m} = \sum_{\bk^{(n)}} \prod_{k\in \bk^{(n)}}\delta_{\sigma(k),k} = \sum_{\bk^{(n)}} j_{\bk^{(n)}}(\sigma).
\en{Eq14}  
When substituted into the output probability distribution of Eq. (\ref{Eq2}), the basic  distinguishability function $j_{\bk^{(n)}}(\sigma)$  imposes the same permutation of bosons in the quantum amplitude and its conjugate, $\sigma_1(k) = \sigma_2(k)$ for $k\in \bk^{(n)}$, i.e., the  bosons from the inputs $k\in \bk^{(n)}$ behave as  completely distinguishable bosons (i.e., as classical particles), and the   output probability  factorizes into a product of those for $N-n$ indistinguishable and $n$ distinguishable bosons.
Similarly, in the case of $J$-function   of  Ref. \cite{R1}, i.e., $J_N(\sigma)=x^{N-C_1(\sigma)}$ the probability of Eq. (\ref{Eq2}) expands as follows (exchanging $n$ and $N-n$, for future convenience)   
\begin{eqnarray}
\label{Eq15}
 &&p^{(J_N)}(l_1\ldots l_N) = \sum_{n=0}^N  x^n(1-x)^{N-n}  \nonumber\\
 && \times \sum_{\bl^{(n)}} \sum_{\bk^{(n)}} p^{(J_N,x=1)}(\bl^{(n)}|\bk^{(n)}) 
    \per |U|^2[\bl^{(N-n)}|\bk^{(N-n)} ],\nonumber\\
     \end{eqnarray}
 where $(\bk^{(n)},\bk^{(N-n)})$ and  $(\bl^{(n)},\bl^{(N-n)})$  are some permutations of  $[N]$,   the  subset  $\bl^{(n)} = (l_{\alpha_1},\ldots, l_{\alpha_n})$ is the set of output ports  of $n$  indistinguishable bosons   ($J_N=1$ for $x=1$), and  $ \per |U|^2[\bl^{(N-n)}|\bk^{(N-n)} ] $  is the classical probability (the matrix permanent of the  matrix $|U_{kl}|^2$), which can be cast  in the form of Eq. (\ref{Eq2}) with $J_0(\sigma) = \delta_{\sigma,I}$  (see Ref. \cite{VS2019} for more details).

The cut-off model with   $J_R $ of Eq. (\ref{Eq7}) does not satisfy the positive definiteness condition formulated in Eq. (\ref{Eq13}), which fact can imply that the approximation with such a distinguishability function does allow  some probabilities to be negative    (note that the normalization condition $J_R (I)=1$ is satisfied, thus the probabilities must sum to $1$).   From  Eqs. (\ref{Eq7}) and  (\ref{Eq14}), by   comparing with  Eq.  (\ref{Eq15}), one   can expand the probability  of  the  cut-off model as follows 
 \begin{eqnarray}
\label{Eq16}
 &&p^{(J_R)}(l_1\ldots l_N) = \sum_{n=0}^N  x^n(1-x)^{N-n}  \nonumber\\
 && \times \sum_{\bl^{(n)}} \sum_{\bk^{(n)}} p^{(J_R,x=1)}(\bl^{(n)}|\bk^{(n)})  
    \per |U|^2[\bl^{(N-n)}|\bk^{(N-n)} ],\nonumber\\
     \end{eqnarray}
where $p^{(J_R,x=1 )}(\bl^{(n)}|\bk^{(n)})  $ is given by Eq. (\ref{Eq2}) with the $J_R$ of Eq. (\ref{Eq7}) for  $x=1$ and may  become  negative. 
Since, by  the expression in Eq. (\ref{Eq2}),  it involves  some mutually dependent diagonals of $U$ (here the  ``diagonal" is a product of  elements of $U$ on distinct  rows and distinct columns), such as  $Z_\sigma = \prod_{k=1}^n U_{\sigma(k),l_k}$ (with $\sigma\in S_n$),  having only up to $n^2$  free parameters instead of $n!$ in an arbitrary $Z_\sigma$,  we need  to find out the  amount of negativity   in   the   subspace  spanned by above diagonals, and not in the entire $n!$-dimensional vector space of $Z_\sigma$.

As $N\to \infty$, for a finite $x=O(1)$ one can estimate the number of computations for  the   direct  sampling from the output distribution in the form of  Eq. (\ref{Eq15}) by observing that the binomial distribution $\binom{N}{n}x^n(1-x)^{N-n}$  becomes sharply concentrated in the small interval of size $O(\sqrt{x(1-x)N})$ centered at $\bar{n} = xN$. Thus   computations of the matrix permanents of the average size $xN$ are required for sampling at an error vanishing as $O(N^{-\frac12})$. Therefore, by similar arguments as in Ref. \cite{Cliffords}, asymptotically in $N$,  the  sampling complexity can be estimated to be the same as that of  the ideal Boson Sampling with  only  $xN$ bosons  (similar observation was used in Ref. \cite{Moy}).   This observation sets the base lime  for the number of computations in an approximate model, such as  Eq. (\ref{Eq16}).  
 \section{ Approximate classical simulation of partially distinguishable bosons    }
\label{sec3}

Let us now recall the main ideas of Ref. \cite{R1} for the efficient approximate classical simulation  of Boson Sampling with partially distinguishable bosons. In the exposition of some of the essential   steps we will follow also Ref. \cite{VS2019}.  They are    as follows. 

\noindent \textit{(I).--} One  considers    the   total variation distance $\mathcal{D}=\frac12\sum |p^{(J_N)}-p^{(J_R)}|$  between the   distributions in Eqs. (\ref{Eq15}) and (\ref{Eq16}) averaged  over the random  multiports $U$ chosen according to the Haar measure. For $M\gg N^2$, up to  a small error,  one can use the Gaussian approximation with independent Gaussian distribution for each $U_{kl}$ \cite{AA} instead.  In the latter case it can be shown that (see derivation in Ref. \cite{VS2019})
\be
\langle \mathcal{D} \rangle_U <   \frac12  \sqrt{1 + \frac{e}{(R+2)!} } \frac{x^{R+1}}{\sqrt{1- x^2}}.
\en{Eq17}  
Having shown that the two probability distributions, one proper and one improper, are close on average, one can then use the Markov inequality in the probability to bound the total variation distance at the cost of some    non-zero probability   of failure \cite{R1}.    

\noindent \textit{(II).--} One  show that the total amount of  computations necessary for obtaining a single  probability for the cut-off model scale as $O(R2^R N^{2R})$ and the small fixed  error requires only $R=O(1)$.   For this goal one can use the expression in Eq. (\ref{Eq2}) substituting the distinguishability function of Eq. (\ref{Eq7}) expressed as a sum over the derangements  $J_R(\sigma) = \sum_{n=0}^R x^n \mathcal{I}(D^{(N)}_n)$, where $\mathcal{I}(D^{(N)}_n)$ is the indicator on the subset $D^{(N)}_n$ of permutations in $S_N$ with $N-n$ fixed points, i.e., the derangements of $n$ elements.   The described expansion reads  (we suppress  the output port indices; see also Ref. \cite{R1})
\begin{eqnarray}
 \label{Eq18} 
&& p^{(J_R)} = \sum_{n=0}^R x^{n} \mathcal{U}(D^{(N)}_n), \nonumber\\
&& \mathcal{U}(D^{(N)}_n) \equiv \sum_{\sigma\in D^{(N)}_n} \sum_{\tau} \prod_{k=1}^N U^*_{\sigma[\tau(k)],l_k}U_{\tau(k),l_k}.
\end{eqnarray}
 By  splitting the   input  $N$ indices   into the $n$  derangements, $\bk^{(n)}$, and $N-n$ fixed points, $\bk^{(N-n)}$,  we can  expand the expression in Eq. (\ref{Eq18}) as follows 
\be
 \mathcal{U}(D^{(N)}_n) = \sum_{\bk^{(n)}} \sum_{\bl^{(n)}}\mathcal{U}(D^{(n)}_n)\per |U|^2(\bk^{(N-n)}|\bl^{(N-n)}),
\en{Eq18A}
where the summation over the partition $(\bl^{(n)},\bl^{(N-n)})$ of the output ports     appears as the result of factoring the  second permutation  in the probability formula  in Eq. (\ref{Eq18})  $\tau = (\tau_1\otimes \tau_2)\mu$ with   $\mu \in \frac{S_N}{S_n\otimes S_{N-n}}$ and   $\tau_1\in S_n$ and $\tau_2\in S_{N-n}$, acting on  $\bl^{(n)}$ and $\bl^{(N-n)}$, respectively. Now,  the first factor $\mathcal{U}(D^{(n)}_n)$  in Eq. (\ref{Eq18A})  contains only the derangements   $D^{(n)}_n\subset S_n$, in  the subset of  indices $\bk^{(n)}$, and the second factor only the fixed points  (a probability of $N-n$   classical particles).    The    derangements   do not represent any probability at all (they appear in  complex conjugate pairs, since the  inverse permutation to a derangements  in $D_n$ is also a derangements in the same subset   $D_{n}$).   They   can be computed using either Ryser or Glynn algorithm (see also section \ref{sec5} below, where such computations are discussed in detail for  another purpose -- checking the negativity), whereas the classical probabilities are estimated to a small relative error  $\epsilon$ by  the probabilistic JSV algorithm \cite{JSV} in the   polynomial   time in $(N-n,1/\epsilon)$. 

\noindent \textit{(III).--} One can sample from the approximate probability distribution Eq. (\ref{Eq16}) by  using the  algorithm similar to that of. Ref. \cite{Cliffords}      if one can compute the  probability  with an acceptable \textit{relative} error, i.e., on the order of the  bound on the total variation distance. However, though the cut-off model does not satisfy the positivity constraint of Eq. (\ref{Eq13}), the negativity is automatically bounded as desired.    Indeed, when   two  distributions, one proper and one improper, are  at some  total variation distance $\epsilon$, the amount of negativity (some of the negative probabilities)  is   bounded by  $2\epsilon$,  by the simple fact that the  total if variation distance is  the maximum of the difference in  probability.     It still  remains to find out the effect of the relative   error introduced by  the probabilistic  JSV algorithm.   In Ref. \cite{R1}  the  average difference in probability $\langle |p^{(J=1)} - p^{(J_R)}|\rangle$  for a  given output is   bounded.  Since the average probability $\langle p^{(J=1)}\rangle = \langle p^{(J)}\rangle$  in a random $U$ is uniform,  the  maximum \textit{relative error} in   a probability, including that   introduced by  the JSV algorithm,  becomes  the \textit{absolute error} on the total variation distance.   Thus the maximum relative error in   probability  becomes  the absolute error on the total variation distance   (similar as for the ideal case of Boson Sampling \cite{AA}; this feature for partially distinguishable bosons  has been  discussed before  in  Ref.  \cite{VS2014}).  

For the discussion below, let us  recall the key points on the    averaging  of  the individual terms in the sum over the derangements, $ \mathcal{U}(D^{(N)}_n)$, in Eq. (\ref{Eq18})  
\be
T(\sigma) \equiv \sum_{\tau} \prod_{k=1}^N U^*_{\sigma[\tau(k)],l_k}U_{\tau(k),l_k},
\en{Eq19}
over the  Haar-random multiport   (in the Gaussian approximation for $M\gg N^2$; below we follow the derivation in    Ref. \cite{VS2019}). We have 
\begin{eqnarray}
\label{Eq20} 
&& \langle T(\sigma)\rangle_U = \delta_{\sigma,I}\frac{N!}{M^N},\nonumber\\
&& \langle T(\sigma)T^*(\pi) \rangle_U = \delta_{\sigma,\pi}\frac{N!}{M^{2N}} \chi(C_1(\sigma)),
\end{eqnarray} 
where $\chi(n) =  n!\sum_{k=0}^n \frac{1}{k!}$. The first line in Eq. (\ref{Eq20}) gives the non-zero  average of a  product of    independent  Gaussian random variables  from  a diagonal of $U$  and that of $U^*$  coinciding only  for  $\sigma=I$. The second  line  in  Eq. (\ref{Eq20}) follows from a result in  Appendix A  of Ref. \cite{VS2014} on the averaging  of  a product of four    diagonals, two  of $U$ and two of $U^*$: 
\begin{eqnarray}
 \label{Eq2014}
&&\langle \prod_{k=1}^N U^*_{\sigma[\tau(k)],l_k}U_{\tau(k),l_k}  U^*_{\sigma^\prime[\tau^\prime(k)],l_k}U_{\tau^\prime(k),l_k}\rangle_U \nonumber\\
&& = \frac{2^{C_1(\pi)}}{M^{2N}} \delta_{\sigma^\prime,\sigma^{-1}}\delta_{\tau^\prime,(\pi\otimes I)\sigma\tau},
\end{eqnarray} 
where  in the second delta-symbol permutation $\pi$  acts on  the  fixed points   and $I$ on the derangements of $\sigma$.  The factor  $N!$ in  Eq. (\ref{Eq20}) accounts for the  sum  over $\tau\in S_N$   in  Eq. (\ref{Eq2014}), whereas  $\chi(n)$ is  the sum over   $\pi\in S_{C_1(\sigma)}$ as follows 
\be
\chi(n) = \sum_{\pi\in S_n} 2^{C_1(\pi)} =  \int\limits_1^\infty \mathrm{d}t e^{1-t}t^n.
\en{chi} 
We have uncorrelated terms $T(\sigma)$ in the summation in Eq. (\ref{Eq18}), where only  the classical term $T(I)$   has a   non-zero average.  The same applies to similar terms in $ \mathcal{U}(D^{(n)}_n)$ in  Eq. (\ref{Eq18A}) (with $N$ substituted by $n$),  with the average  being  always zero.   The bound in Eq. (\ref{Eq17}) follows from the bound on the probability difference     \mbox{$X = |p^{(J_N)} - p^{(J_R)}|$} by its variance  $\langle X\rangle \le \sqrt{ \langle X^2}\rangle$, where the second moment is the variance due to the same average probability.   Moreover,     the square root of the variance  is bounded by the inverse   total number  $\frac{N!}{M^N}$ of  the probabilities  in the output probability distribution (in the no-collision regime).    These  observations allow for  derivation of   the bound in Eq. (\ref{Eq17}).    
  
The   approach  of Ref. \cite{R1}  leads to an   approximate sampling algorithm for partially distinguishable bosons, at least in the model considered in Eq. (\ref{Eq7}).  
 Our main goal below  is  investigate  whether one can do any better.  For instance, can one find any other expansion of the same probability,  alternative  to that of Eq. (\ref{Eq18})  in order to implement sampling  from the classical probabilities in linear time, instead of employing numerically intricate  probabilistic JSV algorithm? 
Such an alternative expansion  will be discussed in  the following section.  
 
The   model  of Eq. (\ref{Eq6}) is also applicable  to the experimental Boson Sampling  with photons  (the most important reason to keep the model for further discussion).  It can serve as  description of the realistic optical  setup, since one has infinite number of free parameters in a photon state (its spectral shape), and the experimental photons  are described by   mixed states. If we assume that the latter are   close with probability $x$ to a pure state $|\phi_0\rangle$ and have a long tail over the orthogonal complement, then in  Eq. (\ref{Eq6})   the orthogonal complementary states $|\phi_i\rangle$, one for each photon,  describe $N$ states selected  at random from the  infinite-dimensional   subspace of the states orthogonal to  $|\phi_0\rangle$.

\section{ Estimating the  negativity   }
\label{sec5}
 
Below  a numerical evidence of  negativity in the quantum factors in the convex sum expansion of the output probability distribution is presented.  There are two types on negativity: the negativity in the approximate probability distribution, reported in Ref. \cite{R1}, and the negativity in a factor in the   convex-sum expression in  Eq. (\ref{Eq16}). The latter negativity  will be numerically estimated below, it prevents  direct  sampling  from   the convex-sum expression  of the approximate probability and forces one to resort to the JSV algorithm for computation of  the classical probabilities.  

The   probability  factor  $ p^{(J_R,x=1 )}(\bl^{(n)}|\bk^{(n)})$  in Eq. (\ref{Eq16}) has    $J_R$ of Eq. (\ref{Eq7})  now for a variable number of bosons $n$   satisfying $n\ge R$ (for $n<R$ the quantum  probability is   positive)  and  $x=1$.  As discussed in section \ref{sec2}, the lack of the positive definiteness of  the distinguishability function   causes some of $ p^{(J_R,x=1 )}(\bl^{(n)}|\bk^{(n)})$    to  become negative.     But how much negativity is there?  Since this depends on the multiport matrix $U$ projection on the negative subspace of the non-positive definite distinguishability function $J_R(\sigma_1\sigma^{-1}_2)$,  considered as the matrix element indexed by  permutations $\sigma_1,\sigma_2$ as in Eqs. (\ref{Eq2}) and (\ref{Eq4}), one can resort to numerical simulations to estimate  negativity  on average in a randomly chosen multiport $U$.  
We only have to  calculate   the probability  for $\bl^{(n)}=\bk^{(n)}= (1,\ldots,n)$  (since we consider a randomly chosen multiport $U$)  
\begin{eqnarray}
 \label{Eq34} 
&& p^{(J_R,x=1 )}  = \sum_{s=0}^R   \mathcal{U}(D^{(n)}_s), \nonumber\\
&& \mathcal{U}(D^{(n)}_s) = \sum_{\sigma\in D^{(n)}_s} \sum_{\tau\in S_n} \prod_{k=1}^n U^*_{\sigma[\tau(k)],k}U_{\tau(k),k}.
\end{eqnarray}

Before  proceeding  to discuss the numerical data, let us  see what one can  expect by trying an analytical analysis. To this goal we can apply the averaging over 
Haar-random multiport $U$ by employing the Gaussian approximation given by  Eqs. (\ref{Eq19})-(\ref{Eq20}) in order to estimate the variance. 
First of all,  let us apply this  approach to the full quantum probability $p^{(J=1)}$ (using  $n$ temporarily as the total number of bosons), obtained by  setting $R=n$ in Eq. (\ref{Eq34}),  i.e., to a positive probability.   We  get the following results (derived before in  Ref. \cite{AA}):
\be 
\langle p^{(J=1)}\rangle_U = \langle \sum_{s=0}^n \mathcal{U}(D^{(n)}_s)  \rangle_U   =  \sum_{s=0}^n\delta_{s,0}\frac{n!}{M^n}  = \frac{n!}{M^n}
\en{AvQp}
and 
\begin{eqnarray}
\label{VarQp}
&&  \langle {p^{(J=1 )}}^2\rangle_U  = \langle \sum_{s,t=0}^n \mathcal{U}(D^{(n)}_s) \mathcal{U}(D^{(n)}_t)\rangle_U  \nonumber\\
&& = \frac{n!}{M^{2n}}\! \sum_{\sigma }\chi(C_1(\sigma))   
  = \frac{n!}{M^{2n}}\int\limits_1^\infty\rd t e^{1-t}\sum_{\sigma }t^{C_1(\sigma)}  \nonumber\\
  && =(n+1)\left(\frac{n!}{M^n}\right)^2, 
\end{eqnarray}
where  the cycle sum over the symmetric group has been used  \cite{Stanley}
\[
\sum_{\sigma\in S_n}t^{C_1(\sigma)} = \left(\frac{\rd}{\rd z}\right)^n_{z=0} \frac{e^{(t-1)z}}{1-z}
\]
to perform  the integration over $t$.  Moreover,  the terms with different  derangements  $\mathcal{U}(D^{(n)}_0),\ldots, \mathcal{U}(D^{(n)}_n)$ are mutually uncorrelated by Eq (\ref{Eq22}) (since the subsets of the respective permutations are non-overlapping),  each contributing to the variance   the square of the average of the classical term (with   $\sigma = I$)  $\langle \mathcal{U}(D^{(n)}_0)\rangle = \frac{n!}{M^n}$. The above calculation illustrates the following point: in the Gaussian approximation  of the Haar-random multiport $U$, the positive by definition probability is a sum of $n+1$ uncorrelated random variables with only the classical term  having non-zero average (equal to the average probability) and the rest with zero average.  The  probability of the partially distinguishable bosons with $J_R$ of Eq. (\ref{Eq7}) is also composed of the same random variables, with, however, one important difference: they are multiplied by the respective powers  of  $x<1$, thus the variances are weighted accordingly.   This observation allows to estimate the average  total amount of negativity in the cut-off model, reported before in Ref. \cite{R1}: it is bounded by  the square root of   the variance of   the  terms subtracted from the full quantum probability, i.e. the terms with  $s=R+1\ldots, n$, multiplied by the respective powers of $x$, i.e., the same    bound  as in  Eq. (\ref{Eq17}).  

Now we can return to  the probability factor given in  Eq. (\ref{Eq34}). In this case  $ x=1$, therefore, the     terms subject to the cut-off   are not weighted by the powers of a small parameter. In this case  it is tempting to conclude that   there is a   finite amount of  negativity. On the other hand, if we apply the same  arguments  to the full quantum probability, we would get the same conclusion.  The error of such conclusion lies in the assumption that uncorrelated terms  contribute independently, but random variables can be uncorrelated and still dependent one on the other (e.g., in a nonlinear way: take $x$ and $x^2$ for a random $x$ with the symmetry $x\to -x$).   Therefore, though the above arguments allow one  to explain a small amount of negativity in the cut-off model due      the  higher  powers of a small parameter $x$, they  do not  allow us to conclude on the negativity in the probability  in Eq. (\ref{Eq34}).  One  must therefore   resort to numerical simulations.  

To numerically compute the  sums involving averaging over the derangements $\sigma$ in Eq. (\ref{Eq34})  we adopt the   method  employed for computation of the matrix permanents \cite{Ryser,Glynn}  (i.e., for averaging  over the whole  symmetric group).  One way is to use   the  inclusion-exclusion principle when averaging over the symmetric group   as  in Ryser's algorithm \cite{Ryser}.  Denoting by $\bl^{(t)}$ the excluded  subset  of size $t$ from  $[n]=\{1,\ldots, n\}$,  we have 
 \begin{eqnarray}
 \label{Eq35}
 && \sum_{\tau\in S_n} \prod_{k=1}^n U^*_{\sigma[\tau(k)],k}U_{\tau(k),k}\nonumber\\
 && = \sum_{t=0}^{n-1} (-1)^t\sum_{\bl^{(t)}}\prod_{k=1}^n \left( \sum_{l\notin \bl^{(t)}}U^*_{\sigma(k),l} U_{kl}\right)
 \end{eqnarray} 
  (for $t=n$  the product of diagonals becomes  empty). Now we need to perform the second summation over the derangements $\sigma\in D^{(n)}_s$ in Eq. (\ref{Eq34}).  To this goal we introduce    a  tensor  function  $W(\xi)$ of  a dummy variable $\xi$ as follows:
 \be
 U^*_{j,l} U_{kl} \rightarrow W_{j,k,l}(\xi) =  \left\{ \begin{array}{cc} U^*_{j,l} U_{kl} ,& j\ne k; \\ \xi |U_{kl}|^2,& j=k. \end{array}\right.
 \en{Eq36}
 With this definition, the average  over the derangements in Eq. (\ref{Eq34}) becomes a Taylor expansion term in $\xi$ of the respective average  over the symmetric group,
 \begin{eqnarray}
 \label{Eq37}
&&  \sum_{\sigma\in D^{(n)}_s}  \prod_{k=1}^n U^*_{\sigma(k),k}U_{k,k} \nonumber\\
&& =  \frac{1}{s!}\left(\frac{d}{d\xi}\right)^s_{\xi=0} \sum_{\sigma\in S_n}\prod_{k=1}^n W_{\sigma(k),k,l}(\xi),
 \end{eqnarray}  
  to which we can apply the same inclusion-exclusion  method as in Eq. (\ref{Eq35}). Finally,  combining the two independent  summations in Eqs. (\ref{Eq35}) and (\ref{Eq37})  and evaluating the   derivative   of the  polynomial function in $\xi$  by  an appropriate    averaging  over the discrete  phase $\xi = e^{i\theta(q)}$, where  $\theta(q) = \frac{2\pi}{n+1}q$ and  $q=0,\ldots, n$, we obtain the final formula for the double sum as follows
  \begin{eqnarray}
  \label{Eq38}
  && \mathcal{U}(D^{(n)}_s) = \frac{1}{n+1}\sum_{q=0}^n e^{-i\theta(q)s} \sum_{t=0}^{n-1} (-1)^t\sum_{\bl^{(t)}} \sum_{f=0}^{n-1} (-1)^f \sum_{\bk^{(f)}}\nonumber\\
 && \prod_{k=1}^n \left(  \sum_{l\notin \bl^{(t)}} \sum_{j\notin\bk^{(f)}} W_{j,k,l}(e^{i\theta(q)})\right).
  \end{eqnarray}
To reduce the number of computations one can perform summation  in Eq. (\ref{Eq38}) of the  exponent over   $s$ from the required set  before summations over the inclusion-exclusion sets.   
The same approach can be used to adopt Glynn's algorithm  \cite{Glynn}  with some advantage of using the recursive computations as in Ref. \cite{Cliffords}.   The above algorithm  requires $O(n^2 4^n)$ computations, where the base $4$ is due to the double summation over the inclusion-exclusion sets.  It allows to compute  the probability distribution over the  Haar-random multiport on a personal computer  for small number of bosons ($n\le 12$).   
 
\begin{figure}[h]
\begin{center}
      \includegraphics[width=.55\textwidth]{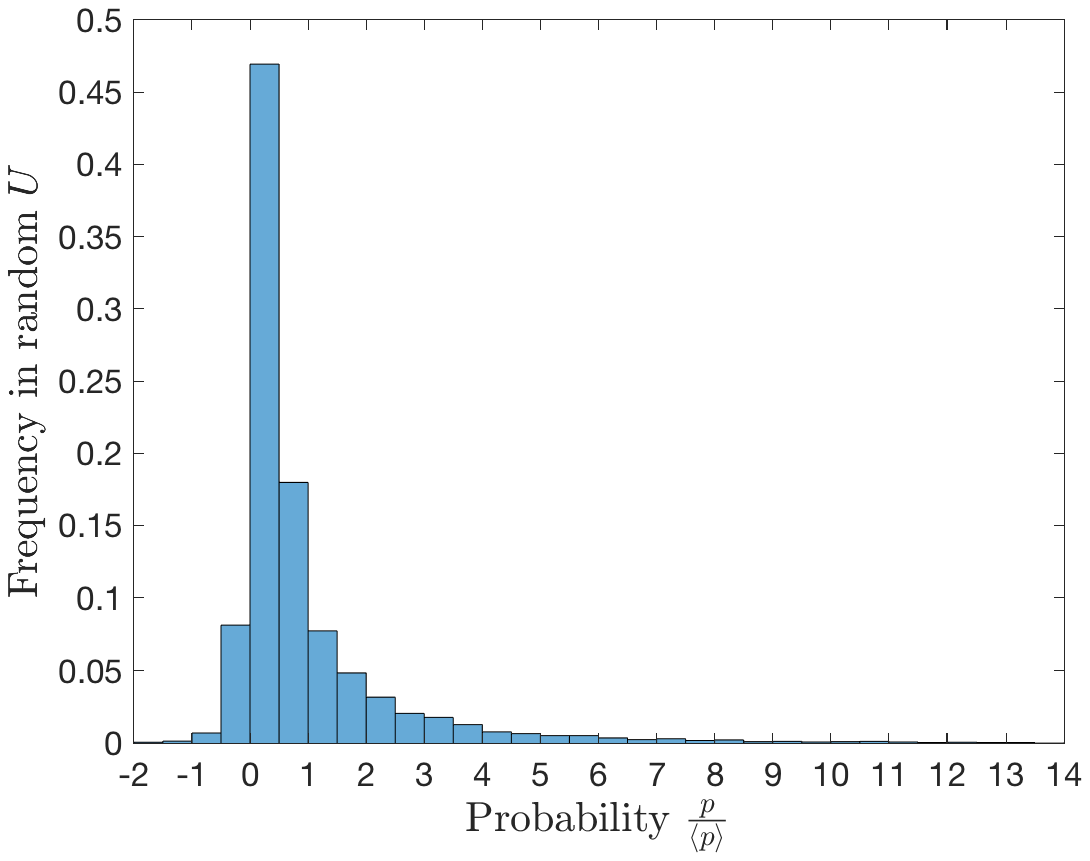} 
     \caption{ Histogram of the   probability in Eq. (\ref{Eq34}) in a random interferometer $U$. Here $n = 7$ and $R = 4$ the histogram is obtained by boxing of the probabilities by their relative occurrence in a randomly chosen multiport from the $5000$ of randomly generated unitary multiports according to  the Haar measure.         \label{F1} }
   \end{center}
\end{figure}

 The results of numerical simulations are presented in Fig. \ref{F1}, where we give  the distribution of the  probability factor  in Eq. (\ref{Eq34}) over the    complex-valued   matrices with independent Gaussian-distributed elements  (in total $ 5000$  matrices were used)    for $n=7$ and $R=4$. Similar distributions in a random multiport was observed for other values of $n$ and various $R<n$.   The odds of having a negative probability    in a random multiport  remain bounded by $\sim10\%$ for all sets of $n$ and $R$ used in the simulations. Since the numerical simulations  are limited to  small $n$, thus one cannot make any definite conclusions on the negativity behavior for  large numbers $n\sim N\gg 1$.   

\section{Positivity of  the cut-off model }
\label{sec4}
 
In section \ref{sec3} we have two equivalent representations of the cut-off model of distinguishability, with the function $J_R$ of Eq. (\ref{Eq7}), one is given in Eqs. (\ref{Eq16}) and the other in Eqs. (\ref{Eq18})-(\ref{Eq18A}).   Let us  find out whether   it admits yet another form, where the derangements are incorporated into some full probabilities   that, at least in the majority of cases (recall that the whole distribution is improper, it allows for negative content on the order of the cut-off error), can be    positive and this fact would  allow one  to avoid usage of  the probabilistic JSV algorithm for the  classical probabilities. Obviously, one cannot break  the derangement  in such a rearrangement (the  cycles  in the independent  cycle decomposition of a permutation  are irreducible), thus the only hope is to break the classical term. Indeed, the latter  is a convex sum,  since the  classical particles pass the multiport $U$ independently.   We will use the following identity  for some  $R\le K\le N$  (recall that $n\le R$) 
\begin{eqnarray}
\label{Eq22}
&&\per |U|^2 (\bk^{(N-n)}|\bl^{(N-n)}) \nonumber\\
&&= \binom{N-n}{K-n}^{-1}\sum_{{\bk^\prime}^{(K-n)}}\sum_{{\bl^\prime}^{(K-n)}}\per|U|^2({\bk^\prime}^{(K-n)}|{\bl^\prime}^{(K-n)})
\nonumber\\
&&\times \per|U|^2({\bk^\prime}^{(N-K)}|{\bl^\prime}^{(N-K)}),
\end{eqnarray} 
(the primed indices  give  partitions  of the corresponding unprimed ones) which is  obtained by expanding the matrix permanent    twice,   on the rows and on the columns. The inverse binomial  compensates for  the   double counting in  choosing a subset of $K-n$ primed indices from    the total $ N-n$ unprimed  ones twice, once for the rows and another time for the columns (the expansion of a matrix permanent either over the rows or over the columns  involves only one such choice).  Inserting the identity of Eq. (\ref{Eq22}) into  Eq. (\ref{Eq18A}) and the result into the probability of Eq. (\ref{Eq18}), identifying  a factor   similar to that in Eq. (\ref{Eq18A}) but now for $K$ bosons,  one obtains the rearranged form for the probability as follows
\begin{eqnarray}
 \label{Eq23} 
 p^{(J_R)} &=& \sum_{\bk^{(K)}} \sum_{\bl^{(K)}}  p^{(J^{(K)}_R)}(\bl^{(K)}|\bk^{(K)})  \nonumber\\
  & &\times \per|U|^2({\bk}^{(N-K)}|{\bl}^{(N-K)}),
\end{eqnarray}
where $(\bk^{(K)},\bk^{(N-K)})$ is a partition of $N$ input and $(\bl^{(K)},\bl^{(N-K)})$ of $N$ output indices (omitted in  $p^{(J_R)} $) and the first factor  reads  
\be
p^{(J^{(K)}_R)}(\bl^{(K)}|\bk^{(K)})  = \sum_{n=0}^R\binom{N-n}{K-n}^{-1} x^n \mathcal{U}(D^{(K)}_n) 
\en{Eq24} 
with  $ \mathcal{U}(D^{(K)}_n)$ still being defined by Eq. (\ref{Eq18A}) now for $K$  bosons instead of $N$.   
The inverse binomial  accounts for the  double counting in the two-stage choosing  of    $n$    from $N$ indices: first  we choose  $K$ indices from $N$   and then   $n $ from $K$, thus  overcounting by    $\binom{N-n}{K-n}$  choices of $K-n$ indices  from the remaining  $N-n$ indices, which are   complementary    to  chosen   $n$ indices:
\be
\binom{N}{n} = \frac{\binom{N}{K}\binom{K}{n}}{\binom{N-n}{K-n}}.
\en{Eq25}

 The physical meaning  of the rearrangement we have performed is as follows. For each $0\le n\le R$ we have split the set of $N-n$  classical particles into two subsets of $K-n$ and $N-K$ particles, the former   is used to  obtain an expression  which can be interpreted at formally as a  probability  of $K$  partially distinguishable  bosons,  Eq. (\ref{Eq24}).      The probability factor in  Eq. (\ref{Eq24}) is not generally positive due to the inverse binomial factor. Observe that the positive-definite    physical  model of partially distinguishable  bosons,  obtained by setting $K=N$  in Eq. (\ref{Eq23})-(\ref{Eq24}),
\be
p^{(J_R)}   = \sum_{n=0}^N x^n \mathcal{U}(D^{(N)}_n), 
\en{Eq26} 
has no such factor.    A condition on the parameter    $x$ is required to restore  the  positive-definiteness. This can be found by  application of the  condition   in Eq. (\ref{Eq13})  to the distinguishability function in Eq. (\ref{Eq24}), in this  case for the total of $K$ bosons (which  is a free parameter,  apart from the condition $K\ge R$). The distinguishability function $J^{(K)}_R$ of Eq. (\ref{Eq24}) reads  (setting $m=K-n$ in Eq. (\ref{Eq24}))
 \begin{eqnarray}
 \label{Eq27}
 && J^{(K)}_R(\sigma) = \sum_{m=K-R}^K \frac{x^{K-m}}{ \binom{N-K+m}{m}}\delta_{C_1(\sigma),m}, \nonumber\\
 && a_m = \left\{ \begin{array}{cc} 0, &  0\le m <K-R; \\  \frac{x^{K-m}}{ \binom{N-K+m}{m}}, &  K-R \le m\le K, \end{array} \right.
 \end{eqnarray}  
 where $a_m$ are the  coefficients  in  the  respective $F(\sigma)$  from Eq. (\ref{Eq10}). 
 For positive definiteness of $J^{(K)}_R$ we have to  ensure   non-negative  coefficients  $b_{K-R},\ldots, b_K$, where 
 \be
b_n \equiv \sum_{m=K-R}^n \binom{n}{m} (-1)^{n-m}\frac{x^{K-m}}{\binom{N-K+m}{m}}
\en{Eq28} 
 (observe that $b_1=\ldots b_{K-R-1}=0$ due to  $a_m=0$ for $m\le K-R-1$).    Eq. (\ref{Eq27}) imposes a very strict condition on possible values of the distinguishability parameter $x$.   Using the rising  factorial notation $n^{(m)} = (n(n+1)\ldots (n+m-1)$ we have  
 \begin{eqnarray}
 \label{Eq29}
 b_n & = &n! x^{K-n} \sum_{m=K-R}^n\frac{(-x)^{n-m}}{(n-m)!} \frac{1}{ (N-K+1)^{(m)}}\nonumber\\
 &=& n! x^{K-n} \sum_{s=0}^{n-K+R}\frac{(-x)^{s}}{s!} \frac{1}{ (N-K+1)^{(n-s)}}.\qquad
 \end{eqnarray}
 For $n \ll \sqrt{N-K}$ we can  approximate the rising factorial as follows 
 \begin{eqnarray}
 \label{Eq30}
 && (N-K+1)^{(n-s)} = (N-K)^{n-s}\prod_{l=1}^{n-s}\left( 1 + \frac{l}{N-K}\right)\nonumber\\
 &&  = (N-K)^{n-s}\left[1+ O\left(\frac{(n-s)^2}{N-K} \right)\right]. 
 \end{eqnarray} 
 The condition can be always  satisfied   for   $R = O(1)$  and large $N$   by demanding that  $R\ll \sqrt{N-K}$ (recall that, as discussed in section \ref{sec3}, a  finite   $R = O(1)$  is required  for a small error in the total variation distance,   whereas $K\ge R$ is still   free parameter).   Substituting the approximation of Eq. (\ref{Eq30}) into Eq. (\ref{Eq29}) we obtain 
 \begin{eqnarray}
  \label{Eq31}
 b_n& = &\frac{n!x^{K-n}} {(N-K)^n} \sum_{s=0}^{n-K+R}\frac{(-[N-K]x)^{s}}{s!} \nonumber\\
 &\times&\left[1+ O\left(\frac{R^2}{N-K} \right)\right],
 \end{eqnarray}
 The   exponential  sum in   Eq. (\ref{Eq31}) has variable upper limit $n-K+R$  and must be always non-negative. Hence,     the  higher powers  of $(N-K)x$ must contribute only a small correction to the sum over the lower powers.  Therefore the  distinguishability function $J^{(K)}_R(\sigma)$  in Eq. (\ref{Eq24}) becomes positive definite   for the distinguishability parameter
 \be
 x \le \frac{1}{N-K}.
 \en{Eq32}
If one wants to avoid using the JSV algorithm, then  at most  $K = O(\ln N)$ can be used  in Eq. (\ref{Eq32}) for efficient   computation of  the  probability in Eq. (\ref{Eq24})  bypassing the use of the  JSV algorithm, which requires  an exponential in  $K$ number of  computations for $K$ particles   (by the algorithm  of section \ref{sec5} the number of computations  becomes  $O(K^24^{K})$).      Then  Eq. (\ref{Eq32})   becomes  too restrictive on the distinguishability parameter $x$,  as it does not  apply to a finite  $x$  as $N$ scales up.  Moreover,  the    asymptotically average  total number   of bosons $\bar{n}= xN$, as  discussed in section \ref{sec3},   becomes bounded as $N\to \infty$ for  $x$ from Eq. (\ref{Eq32}).     Therefore,  this approach fails to produce any advantage.   

\section{Partial distinguishability  and matrix immanants  }
\label{sec6}

The purpose of this section is to show that negativity   can be avoided in approximations to the noisy Boson Sampling distribution. However, the new approach  demands  development of  the asymptotic character theory for   the symmetric group $S_N$  as $N\to \infty$ (e.g.,   an explicit expression  for the character table valid to a vanishing error as $N\to\infty$).  

  The   distinguishability function  $J(\sigma) = x^{N-C_1(\sigma)}$  besides being  experimentally relevant, as discussed at the end of section \ref{sec3}, happens to be  a class function, i.e., it satisfies the property 
\be
F(\tau\sigma\tau^{-1}) = F(\sigma),
\en{Eq43}
for all $\sigma, \tau \in S_N$, since the number of fixed points remains invariant under the similarity transform (the conjugation in $S_N$) $C_1(\tau\sigma\tau^{-1}) = C_1(\sigma)$. 
This fact allows  one to expand such  $J(\sigma)$  over the irreducible characters $\chi_j(\sigma)$ of   $S_N$, which are themselves some positive-definite (though not normalized) functions on $S_N$ (see, for instance, Ref.  \cite{Weyl}).  This approach is not equivalent to  the standard application of the Schur-Weyl dyality to the action of the group of unitary transformations  on the tensor product of the single-boson Hilbert spaces, employed  for multi-boson interference of partially distinguishable bosons  in Refs.  \cite{imm,CircMod}, as here  the character  theory is  applied to the distinguishability function as an element of the  linear space of  class functions on $S_N$. 

We  will use  the following  facts  (see e.g., Ref. \cite{Weyl}). First,  a character is a trace of a representation of the group, a positive-definite function of a conjugacy class of $\sigma$  i.e.,  all  permutations of the form  $\tau\sigma\tau^{-1}$ for $\tau\in S_N$. Second, an arbitrary character of the group    can be written as a convex sum (precisely, with some non-negative  integer coefficients) over  the orthogonal basis of the  irreducible characters $\chi_j(\sigma)$, where there are so many irreducible characters as the conjugacy classes in $S_N$. 

Let us   derive the expansion of the distinguishability  function $J(\sigma) = x^{N-C_1(\sigma)}$  as a (convex)  sum of the irreducible characters of the symmetric group $S_N$.  To this goal, recall that our  $J(\sigma)$ has the form of  Eq. (\ref{Eq3}), where $\hat{\rho}^{(N)}= \hat{\rho}_1\otimes \ldots \otimes \hat{\rho}_N$ with $\hat{\rho}_k$ from Eq. (\ref{Eq6}). Therefore, it can be also cast as follows 
\be
J(\sigma) = \sum_{n=0}^Nx^n(1-x)^{N-n} \mathrm{Tr}_{\cL_n}\{\hat{P}_\sigma\} ,
\en{Eq44}
where we have introduced  $N+1$ orthogonal  linear subspaces in $H^{\otimes N}$, where  subspace  $\cL_n$   is some  linear  span of $\binom{N}{n}$    orthogonal 
states generated by the action of the   permutations   $\mu \in \frac{S_N}{{S_n}\otimes S_{N-n}}$. In other words,  $\mu$ is the unique  factor in the   decomposition $\sigma = (\sigma_n\otimes\sigma_{N-n})\mu$ where    $\mu$ selects the first $n$ elements of $N$ and $(\sigma_{n},\sigma_{N-n}) $  permutes inside the  two subsets of sizes $n$ and $N-n$. Precisely, subspace $\cL_n$ is generated by the unitary operators  $\hat{P}_\mu$  acting   on the base state $|\Psi_n\rangle\equiv |0\rangle^{\otimes n}\otimes |1\rangle\otimes \ldots \otimes |N-n\rangle $  composed of some orthogonal states $|i\rangle\in H$:  $\langle i|j\rangle = \delta_{ij}$, $i,j = 0,1,\ldots N$. We have 
\begin{eqnarray}
\label{Eq45}
&&\cL_n \equiv \mathrm{Span}\left\{|\mu,n\rangle ;\quad \forall \mu \in \frac{S_N}{S_n\otimes S_{N-n}} \right\}, \nonumber\\
&& |\mu,n\rangle \equiv \hat{P}_\mu|0\rangle^{\otimes n}\otimes |1\rangle\otimes \ldots \otimes |N-n\rangle.
\end{eqnarray}
Taking the  trace over the subspace $\cL_n$   is equivalent to performing summation  of   the average values  of the  operators  $\hat{P}_{\mu^{-1}\sigma\mu}$  on the base state in $\cL_n$,  i.e.,  the trace is the projection on the   permutations $\mu^{-1}\sigma\mu$ with at least  $ N-n$ fixed points: $n+1,\ldots, N$ (otherwise  we get zero). Denoting $\bk^{(N-n)}=(\mu(n+1),\ldots, \mu(N))$ and the summation over $\mu$ by that over $\bk^{(N-n)}$ we get by using the relation in Eq. (\ref{Eq14}):
\begin{eqnarray} 
\label{Eq46}
&&\mathrm{Tr}_{\cL_n}\{\hat{P}_\sigma\}  =   \sum_{\bk^{(N-n)}} \prod_{k\in \bk^{(N-n)}}\delta_{\sigma(k),k}   \nonumber\\
 &&=  \sum_{m=N-n}^N\binom{m}{N-n} \delta_{C_1(\sigma),m}.
\end{eqnarray}
The above trace is of the matrix   $M_{\mu,\mu^\prime}(\sigma) =\langle\Psi_n| \hat{P}^\dag_\mu\hat{P}_\sigma\hat{P}_{\mu^\prime}|\Psi_n\rangle$, i.e., the  matrix of a  linear  representation of the symmetric group in $\cL_n$, therefore, according to the general theory of   group characters,   the function in Eq. (\ref{Eq46})  is   the corresponding  character of the symmetric group (generally reducible).    By this fact,  we must  have 
  \be
 \mathrm{Tr}_{\cL_n}\{\hat{P}_\sigma\}  = \sum_{j} \mathcal{M}_{n,j}\chi_j(\sigma),\quad \mathcal{M}_{n,j} \in \mathcal{N}_0.
 \en{Eq47}
 where integer $\mathcal{M}_{nj}$ counts the number of irreducible representations with the character $\chi_j$ in the decomposition of the representation in the subspace $\cL_n$. Two irreducible characters are well-known: the trivial one $\chi_1(\sigma) =1$ ($J$-function for the indistinguishable bosons) and the sign character $\chi_2(\sigma) = \mathrm{sgn}(\sigma)$ (that for the indistinguishable fermions), whereas all other correspond to the so-called matrix immanants  \cite{MLA} .  
 
The above suggest the  main idea: to consider the expansion resulting from Eq. (\ref{Eq44}) and (\ref{Eq47})
\begin{eqnarray}
\label{Eq48}
&& J(\sigma) = \sum_{j} \chi_j(\sigma)\sum_{n=0}^N \mathcal{M}_{n,j} x^n(1-x)^{N-n} \nonumber\\
&& = \sum_j q_j(x) \frac{\chi_j(\sigma)}{\chi_j(I)}, 
\end{eqnarray}
i.e., a convex sum  ($0\le q_j(x)\le 1$, $\sum_j q_j(x) = 1$) over  the  positive-definite normalized irreducible characters, satisfying all the properties of a distinguishability function, as discussed in section \ref{sec2},  generalizing the concept of quantum particles beyond bosons and fermions.   The   behavior of the  coefficients $q_j(x)$ as functions of the distinguishability parameter $x$  could then suggest   an approximation which  would  result in  a small total variation distance error between the distributions and, at the same time, retain the  positive-definiteness property of the proper  distinguishability function. 

The outcome of the above idea is far from clear from the outset, since even the distinguishability function of  classical particles $J_{class}(\sigma) = \delta_{\sigma,I}$  can be expanded as in Eq. (\ref{Eq48}) over \textit{all} the irreducible characters, including  completely  indistinguishable  bosons ($\chi_1(\sigma)$), though one can sample from the classical particles linearly in the total number of them. On the other hand, asymptotically, as    $N\to \infty$, it may happen that the contribution   from the individual characters becomes exponentially small, due to large    number of classes in the symmetric group $S_N$, coinciding with the partition function $P(N)\sim e^{\pi\sqrt{2N/3}}$.

The above program  necessitates knowledge of the  asymptotic values of the irreducible  characters of the symmetric group $S_N$ as $N\to \infty$,  i.e., a   workable formula instead of  an algorithm  for the   asymptotic  character tables (their values on the conjugacy classes). Moreover, it  necessitates  the asymptotic complexity of all the matrix immanants, the subject under intensive  investigation (for  a recent  review, see Ref. \cite{immCo1}).  Given these two mathematical problems solved,  one could use the results in the search for a better classical algorithm for noisy Boson Sampling, where uncontrollable partial distinguishability of bosons serves as   noise.  There are also equivalence relations  between different models of noise in Boson Sampling \cite{VS2019}, which could allow   general conclusions on the effect of noise. 


\section{Conclusion}
\label{sec7}
The main goal of the present work has been to find a better algorithm for sampling from a noisy Boson Sampling distribution, with the partial distinguishability of bosons serving as noise, which does not require  to  compute the  probabilities of multiple transitions of   completely distinguishable bosons.  All attempts  to find such a better algorithm seem to be  faced with the negativity in the approximating  distribution, if the latter is obtained by imposing a cut-off on the order of multi-boson interferences. The negativity in the approximating distribution stems from the lost positive-definiteness of the  respective   partial distinguishability function.  Numerical evidence points on  a finite amount of negativity, at least for small numbers of bosons, accessible to numerical simulations.  Rewriting  the approximate   (i.e., truncated)  probability distribution  for $N$ bosons in total in another, physically more clear, form, as a  convex sum expansion of terms corresponding  each to a smaller total number of bosons $K$, which would include as a subset the bosons participating in the  lower-order interferences accounted for by the approximation,   does not help as in this new form each term  with $K$ bosons  may still   be a non-positive probability,  for  the distinguishability parameter  satisfying $x\le \frac{1}{N-K}$, whereas  the computational complexity scaling  as $O(K^24^K)$ by either the modified Ryser or Glynn algorithms.  This results in a too narrow region of the distinguishability parameter $x$, vanishing  as $(N- O(\ln N))^{-1}$, if we aim to have an algorithm for approximate sampling asymptotically polynomial in $N$. 

The judgement that there must always be negativity in  the approximate efficient classical sampling from a noisy Boson Sampling is, nevertheless,  premature. A better sampling algorithm, devoid of  the negative probabilities and, hence, of  computations of the probabilities of joint  transitions of completely distinguishable bosons,     might still  be  possible.  I  have outlined the direction for  a new program which would  supply only  the  proper   approximate distributions close to that of  noisy Boson Sampling. To pursue this program, however,  the asymptotic character theory of the symmetric group has to be developed first, which sound as  a project in its own right. Moreover, the full picture of the asymptotic computational complexity of the so-called matrix immanants  is  required for estimating the computational complexity of the approximating distributions.   

 \medskip
\section{Acknowledgements}  
This work  was supported by the National Council for Scientific and Technological Development (CNPq) of Brazil,  Grant 307813/2019-3.



\begin{thebibliography}{99}

  \bibitem{AA} S. Aaronson and A. Arkhipov, Theory of Computing \textbf{9},  143 (2013).
 
 
 \bibitem{QSup} A. W. Harrow and A. Montanaro,  Nature \textbf{549}, 203  (2017).
 
 
\bibitem{20Ph60M}  H. Wang, J. Qin, X. Ding, M.-C. Chen, S. Chen, X. You, Y.-M. He, X. Jiang, L. You, Z. Wang, C. Schneider, J. J. Renema, S. H\"ofling, C.-Y. Lu, J.-W. Pan,  Phys. Rev. Lett. \textbf{123}, 250503 (2019).


\bibitem{QSBS} A. Neville, C. Sparrow, R. Clifford, E. Johnston, P. M. Birchall, A. Montanaro, A. Laing,
Nature Physics \textbf{ 13}, 1153 (2017).
 
\bibitem {Cliffords} P. Clifford, and R. Clifford,  arXiv:1706.01260 (2017).

  \bibitem{GBS1}  A. P.  Lund, A. Laing, S. Rahimi-Keshari, T. Rudolph, J. L.  O'Brien, and T. C. Ralph, Phys. Rev. Lett. \textbf{113}, 100502 (2014).
 
 \bibitem{GBS2}   C. S. Hamilton, R. Kruse, L. Sansoni, S. Barkhofen, C. Silberhorn, and I. Jex,   Phys. Rev. Lett.  \textbf{119}, 170501 (2017).

 
 \bibitem{ExpGBS2}  H.-S. Zhong \textit{et al},  Quantum computational advantage using photons, Science \textbf{370},   1460 (2020). 
 
 \bibitem{R1}   J. J. Renema, A. Menssen, W. R. Clements, G. Triginer, W. S. Kolthammer, and I. A. Walmsley,
 Phys. Rev. Lett. \textbf{120}, 220502  (2018). 

 \bibitem{LP} A. Leverrier and R. Garc{\'i}a-Patr{\'o}n,  	Quant. Inf. \& Computation  \textbf{15}, 0489 (2015). 


\bibitem{KK}  G. Kalai and G. Kindler,  	arXiv:1409.3093 [quant-ph]. 

 \bibitem{Arkh}  A. Arkhipov,  Phys. Rev.  A  \textbf{92}, 062326 (2015). 


\bibitem{JSV} M. Jerrum,  A.  Sinclair, and E. Vigoda, Journal of the ACM \textbf{51}, 671 (2004).



\bibitem{Moy} A. E. Moylett,  R. Garc\'ia-Patr\'on, J. J. Renema, and P. S. Turner.  Quantum Sci. Technol.,  \textbf{5},  015001  (2020).

\bibitem{VS2014}      V. S. Shchesnovich, Phys. Rev.  A  \textbf{89},   022333  (2014).

\bibitem{PartDist} V. S. Shchesnovich, 	 Phys. Rev. A  \textbf{91}, 013844  (2015).

\bibitem{Minc} H. Minc, \textit{Permanents, Encyclopedia of Mathematics and Its Applications}, Vol. \textbf{6} (Addison-Wesley Publ. Co., Reading, Mass., 1978).


\bibitem{PRL2016} V. S. Shchesnovich, Phys. Rev. Lett.   \textbf{116}, 123601 (2016).

\bibitem{VS2019} V. S. Shchesnovich, Phys. Rev. A  \textbf{100}, 012340 (2019). 



  
\bibitem{Ryser} H. Ryser, \textit{Combinatorial Mathematics}, Carus Mathematical Monograph No. 14. (Wiley, 1963).


\bibitem{Glynn} D. G. Glynn, Eur. J. of Combinatorics \textbf{31}, 1887 (2010). 


  



\bibitem{VS2015} V. S. Shchesnovich, Phys. Rev. A  \textbf{91}, 063842 (2015).

\bibitem{Brod} S. Aaronson and D. J. Brod,   Phys. Rev.  A  \textbf{93}, 012335 (2016).


 



  

 \bibitem{Stanley}   R. P. Stanley, \textit{Enumerative Combinatorics}, 2nd ed., Vol. 1 (Cambridge University Press, 2011). 
 
 \bibitem{Weyl} H. Weyl, \textit{The Classical Groups: Their Invariants and Representations} (Princeton University Press; 2nd   ed.   1997). 
 
 \bibitem{imm}  M. Tillmann, S.-H. Tan, S. E. Stoeckl, B. C. Sanders, H. de Guise, R. Heilmann, S. Nolte, A. Szameit, and P. Walther, 
  Phys. Rev. X \textbf{5}, 041015 (2015).
  
\bibitem{CircMod} A. E. Moylett and P. S. Turner, Phys. Rev. A \textbf{97}, 062329  (2018). 

 \bibitem{MLA} R. Merris, \textit{Multilinear algebra} (CRC Press, 1997).
 
    
 \bibitem{immCo1}  R. Curticapean, 	arXiv:2102.04340 [cs.CC]. 

 \end{thebibliography}
\end{document}